# Phase-coexistence and glass-like behavior in magnetic and dielectric solids with long range order

S. B. Roy and P. Chaddah
Magnetic and Superconducting Materials Section,
Materials & Advanced Accelerator Sciences Division
Raja Ramanna Centre for Advanced Technology, Indore – 452013, India

## Abstract

Phase-coexistence in the manganese-oxide compounds or manganites with colossal magneto-resistance (CMR) has been generally considered to be an inhomogeneous ground state. An alternative explanation of phase-coexistence as the manifestation of a disorder-broadened first order magnetic phase transition being interrupted by the glasslike arrest of kinetics is now gradually gaining ground. This kinetic arrest of a first order phase transitions, between two states with long-range magnetic order, has actually been observed in various other classes of magnetic materials in addition to the CMR manganites. The underlying common features of this kinetic arrest of a first order phase transition are discussed in terms of the phenomenology of glasses. The possible manifestations of such glass-like arrested states across disorder-influenced first order phase transitions in dielectric solids and in multiferroic materials, are also discussed.

# 1 Introduction

First order phase transitions are defined by a discontinuous change in entropy at the transition temperature $T_C$ (resulting in a latent heat) and a discontinuous change in either volume or magnetization (depending on whether $T_C$ changes with pressure or with magnetic field). Since liquids and solids have different densities, the crystallization of a liquid entails motion at a molecular level that requires non-zero time. The concept of rapidly freezing a liquid out of equilibrium has been exploited for producing splat-cooled metallic glasses [1]. As was noted by Greer [1], the glass would have a density close to that of the liquid.

We have noted recently that the kinetics of a first order phase transition is dictated by the time required for the latent heat to be extracted and, in addition to the well-known quenched metallic glasses, the kinetics of any first order phase transition (including the one between two long-range-order phases) could thus be arrested [2]. We have studied many first order magnetic transitions whose kinetics has been arrested [3-14], and shall point out similar behavior in some published reports on first order dielectric transitions.

While temperature is one thermodynamic variable common to all these classes of first order phase transitions (viz. structural, magnetic, dielectric), the second thermodynamic variable would be pressure, magnetic field, and electric field, respectively. There has been a resurgence of interest in first order phase transitions with magnetic field induced transitions providing the impetus, because of possible applications envisaged for magneto-caloric materials, for materials showing large magneto-resistance, for magnetic shape memory alloys, etc [2]. Unlike pressure, magnetic field does not require a medium for its application and controlling H thus does not complicate controlling temperature. This has enabled some very interesting observations on glasslike metastable states. While electric field has not yet been used extensively as a thermodynamic variable, the scenario is bound to change as more multiferroic materials are discovered.

The role of the second thermodynamic variable in glass formation has been recognized even in structural glasses because the freezing point shifts with pressure, but was exploited only a few years back when elemental germanium was vitrified under pressure, and the glassy state was retained on release of pressure [15]. The use of magnetic field as a second thermodynamic parameter for arresting a first order magnetic transition was conceived and exploited by us over ten years back [3-14]. If the first-order transition occurs over a range of temperatures (as happens in the presence of disorder, or when the transition is accompanied by large strain) and the glass formation temperature ($T_K$) falls within this range, then there is an interesting possibility that the transition is arrested while it is still partial and incomplete. We observed first order magnetic transitions being arrested at low temperatures while still incomplete, resulting in the two magnetic phases coexisting down to the lowest temperature.

First reported by us in the case of a ferromagnetic to anti-ferromagnetic transition in $CeFe_2$-pseudobinary alloys, such persistent phase coexistence during cooling has now been reported across first order magnetic transitions in various classes of magnetic materials including CMR manganites [2]. We shall discuss our studies following various thermomagnetic history paths where the cooling field and the warming field are different, that allow study of reentrant transitions, analogous to devitrification of glass followed by melting of the crystalline state.

## 2 Kinetic arrest of disorder broadened first order magneto-structural transition: Magnetic-glass

A first order magnetic/magneto-structural phase transition across a phase transition temperature $T_N(H)$ line in the two parameter magnetic field (H) - temperature (T) phase space (see Fig.1) of a solid magnetic material is identified by a discontinuous change in entropy (i.e. measureable latent heat) or a discontinuous change in magnetization (M) as this $T_N(H)$ line is crossed by varying either of the control variables T or H. The high- and low-temperature phases coexist at the transition temperature $T_N$, and in the absence of any energy fluctuations the high temperature phase continues to exist as a supercooled metastable state until the temperature $T^*$, which represents the limit of supercooling [16]. A limit of metastability is similarly possible while heating at $T^{**}(H) > T_N(H)$, and this is not shown in the schematic (H,T) phase diagram (Fig.1) for the sake of clarity. Such limits of metastability $H^*$ and $H^{**}$ can also be defined across an isothermal magnetic field induced first order phase transition. In many real magnetic materials instead of latent heat (which is often difficult to determine experimentally) such phase-coexistence and metastability can be used as characteristic observables to identify a first order phase transition [17].

The actual composition of multi-component magnetic alloys, intermetallic compounds and metal-oxide compounds varies around some average composition due to disorder that is frozen in as the solid crystallizes from the melt. In a pioneering work Imry and Wortis [18] showed that such static, quenched-in, purely statistical compositional disorder could broaden a first order phase transition. The associated latent heat could be significantly diminished or completely wiped out by the disorder. This disorder induced broadening of first order phase transition can take place for sufficiently large range of disorder correlations, greater than the order-parameter correlation length and the length defined by the ratio of the inter-phase surface tension and the latent heat [18,19]. This phenomenon of disorder influenced first order phase transition drew significant attention of the theorists [19,20], and its possible role in the interesting properties of CMR manganites was highlighted by detailed computational studies [21]. On the experimental front, finding of an intrinsic disorder-induced landscape of vortex solid melting transition temperatures/fields in a high temperature superconductor BiSrCaCuO [22] had shown the applicability of such a concept in real materials.

It is well known that some liquids, called 'glass-formers', experience a viscous retardation of crystallization during first order phase transition in their supercooled state [23]. Such supercooled liquid ceases to be ergodic within the experimental timescale, and enters a non-equilibrium state called a 'glassy state'. Glass can be considered within a dynamical framework, as a liquid where the atomic or molecular motions are arrested [24], which leads to the general definition of conventional glasses that 'glass is a noncrystalline solid material which yields a broad nearly featureless diffraction pattern'. There is another widely acceptable picture of glass as a liquid where the atomic or molecular motions (or kinetics) are arrested. Within this latter dynamical framework, 'glass is time held still' [24]. Various dynamical features typically associated with glass formation are not necessarily restricted to materials with positional disorder [24].

It is reasonable to assume that during glass formation the viscous retardation of crystallization on an experimental timescale would occur below some temperature $T_K$. In the case of metallic glasses $T_K < T^*$ the limit of supercooling of the liquid, hence the state in which the glass can be formed is unstable. In such cases the rapid quenching of the liquid is essential so that the kinetics is arrested before it can effect a structural change. On the other hand in the case of $T_K > T^*$ the glass is formed in a state that is metastable (in terms of free energy), and the glass formation can be caused by slow cooling, as in the well known glass-former O-terphenyl [23]. In this case the system is already trapped in a deep valley in the potential energy landscape [23] corresponding to a glass structure, although it has reached the spinodal point in the free energy configuration. The potential energy minimum corresponding to a crystalline structure is well separated from this deep valley and a glassy state does not get transformed into a crystalline state with finite energy fluctuations within a finite experimental timescale.

As shown in the schematic in Fig. 1, $T_N$ drops with increasing H, just like the transition temperature ($T_M$) drops with increasing pressure whenever a liquid expands on freezing. Within the realm of experimentally achievable steady pressures $T_M$ drops by 25% or less, whereas in many ferromagnetic (FM) to antiferromagnetic (AFM) transitions, $T_N$ drops almost by 100% with experimentally achievable steady magnetic fields. As the drop in the transition temperature slows the kinetics of the phase transition, it is more likely to produce a glass-like arrested state. We thus see the possibility of going from the scenario of $T_N \gg T^*$ to $T_N < T^*$ with experimentally achievable magnetic fields, or of going from a metglass like situation (which requires rapid quenching from liquid state to avoid crystallization) to an O-terphenyl ( a standard glass-former) like situation in the same material. This glass-like arrested state formed by arresting a first order magnetic phase transition is referred to as a magnetic-glass, and shows various attributes (including time dependent behavior) similar to structural glasses. Indeed it has recently been observed that in certain regions of the H–T phase diagrams of the different classes of magnetic materials (CeFe$_2$ based alloys, Gd$_5$Ge$_4$, CMR mangnites, Ni–Mn–In and Ni–Mn–Sn Heusler alloys and Fe–Rh ) there is a viscous retardation of nucleation and growth of the low temperature phase across a magneto-structural first order phase transition leading to such a magnetic-glass state [3-14] . In all these materials quenched disorder broadens the first order phase transition as envisaged within the Imry-Wortis framework [18], and there is a landscape or band of transition temperature $T_N$ (H) or field $H_M$ (T) and limit of supercooling (or superheating) temperature $T^*$(H) (($T^{**}$(H))) or field $H^*$(T) (($H^{**}$(T))) [25] instead of a single transition temperature and limit of supercooling (superheating). The schematic in Fig.2 shows disorder broadening of $T_N$, as well as of $T^*$ and $T^{**}$, because they are now defined over regions of the length-scale of the correlation length and have values dictated by local composition [26].

The schematic in Fig.3 shows how the transition proceeds over a range of temperatures for a fixed value of H. For an ultrapure material the material can be taken to the limit of superheating (supercooling) $T^{**}$ ($T^*$) without any nucleation of the high (low) temperature phase and the phase transition onset temperature $T_1$ ($T_3$) coincides with the $T^{**}$ ($T^*$) (see Fig. 3(a)). In the presence of small disorder, nucleation and growth of the high (low) temperature phase starts before reaching $T^{**}$ ($T^*$), and we denote the new positions of $T_1$ ($T_3$) by $T_{1a}$ ($T_{3a}$) . In many materials of practical interest, the quenched-disorder will thus give rise to $T^{**}$ ($T^*$) bands, corresponding to the range between start and completion of phase transition while cooling $T_{3a}$ and $T^*$ (and start and completion of phase transition while heating $T_{1a}$ and $T^{**}$ respectively)

[26]. Fig. 3(b) shows the case when the band becomes very broad, and $T_3$ is shifted to a position which is higher than $T_1$. This schematic in Fig. 3(b) corresponds to the extensively studied Magnetic Shape Memory alloys (MSMA) where glasslike kinetic arrest of the first order martensitic transition has been reported. We must point out that this schematic visually emphasizes that the underlying thermodynamic transition temperature cannot be estimated as Tm=[Ms +Af]/2 [27]; it is best approximated by [Ms+Mf+As+Af]/4 [28]. Further, we now have the very interesting possibility that $T_K$ intersects the T* band over some range of H. In this range of H the broadened transition is initiated on cooling and proceeds, but is interrupted (or arrested) midway as the kinetics gets arrested. As stated above, T* is decreasing with rising H. Assuming that $T_K$, where the arrest (or interruption) occurs remains same as H rises, it follows that a smaller fraction of the transition has taken place when the transition is interrupted. One manifestation of this is that thermomagnetic irreversibility associated with zero field cooled(ZFC)/field cooled(FC) magnetization measurements would increase with rising H. This is in striking contrast with thermomagnetic irreversibilities associated with pinning of domain walls, of easy axis and spin-flip process, or even with spin-glasses. A clear cut manifestation of this contrasting behavior has been reported by Sharma et al [7] in the case of NiMnIn shape memory alloy. Figure 4, presents the temperature dependence of magnetization of this NiMnIn alloy obtained under zero field cooled (ZFC), field cooled cooling (FCC) and field cooled warming (FCW) protocol. The austenite to martensite phase transition is accompanied by a sharp drop in magnetization [7]. The thermal hysteresis in the FCC/FCW magnetization across the austenite-martensite transition is associated with the first order nature of this phase transition. In addition there is a distinct thermomagnetic irreversibility between the ZFC/FC magnetization in the matertensitic phase at the low applied filed H = 100 Oe, which gets reduced with the increase in H and becomes completely reversible for H = 10 kOe (see Fig.4). This hysteresis is attributed with the hindrance of domain wall motion in the ferromagnetic martensitic phase. However, in applied H beyond 10 kOe, the ZFC/FC thermomagnetc hysteresis reappears and then rises with further increase in H. This hysteresis is due to the glass-like kinetic arrest of the first order austenite-martensite phase transition in the presence of high H. The non-equilibrium nature of the magnetic-glass state is evidenced by the fragility (devitrification) of this state by multiple temperature cycling (see the inset of the upper figure in 4(b)). It has been further observed that in a Fe-doped NiMnIn Heusler alloy the kinetic arrest of the austenite-martensite phase transition takes place even in the absence of an applied magnetic field [29].

$T_K$ is also influenced by disorder; it can also have different values (even for a specific cooling rate) for different regions having the length-scale of the correlation length. Schematic-4 presented in Fig. 5 shows this interesting scenario; the glasslike arrested state can now undergo only partial devitrification when H is varied at some temperature like (see point F in Fig.5). This gives rise to an anomalous behavior in the form of the virgin magnetization or resistivity curve lying outside the envelope curves as H is varied isothermally [25]. This anomalous behaviour was first reported in Al-doped $CeFe_2$ in both resistivity measurements and in magnetization measurements (see Fig. 6 and 7). With increasing H the AFM state undergoes a broad first order phase transition to the FM state as it crosses the superheating band, with the AFM state not being fully recovered even when H is reduced to below the supercooling T** band because it has only partially crossed the $T_K$ band. The remaining AFM state can be recovered by heating, which also causes devitrification. We would like to assert that it was shown in these pioneering works that de-arrest was caused both by raising T and by lowering H. As we shall discuss below, in materials like $Gd_5Ge_4$ having an AFM-to-FM transition with lowering temperature, isothermal increase of H at the lowest temperature ( say T = 2K) causes devitrification of the kinetically arrested first order magnetic transition. In the light of these observations, we believe any rechristening of "kinetic arrest" as "thermal transformation arrest" [30] would belie the underlying physics of the phenomenon.

In the case of materials like $Gd_5Ge_4$ that have an AFM-to-FM transition with lowering T, the transition temperature $T_N$ is lowered by decreasing H. Cooling in higher fields inhibits the process of "kinetic arrest", and less of the transition has proceeded before it is arrested (or interrupted) [6]. In the isothermal variation of H at low T, the glasslike arrested state is de-arrested with increasing H [6]. The virgin ZFC state is the arrested AFM state, while the remnant zero-field state after isothermal cycling of H is the equilibrium FM state [6]. This kinetic arrest of the AFM-FM transition in $Gd_5Ge_4$ has actually been visualized with micro-Hall probe scanning study [31]. Further the devitrification process of this arrested state has also been studied in details [32]. In Figure 8 we show the T*, T**, and $T_K$ bands for both cases of AFM ground state and of the FM ground state (with the simplifying assumption that the slopes of these bands have no H-dependence) [33]. Even though the ground states are very different, the manifestation of kinetic arrest in isothermal M-H curves is very similar. This behavior is exemplified in Fig.9 with very similar anomalous isothermal virgin M-H curves obtained in $Pr_{0.5}Ca_{0.5}Mn_{0.975}Al_{0.025}O_3$ and $La_{0.5}Ca_{0.5}MnO_3$ with ferromagnetic and antiferromagnetic ground state, respectively [10]. A magnetic-glass state arising out of a kinetic arrest of a first order AFM (charge ordered) to FM transition has been reported in La5/8−yPryCa3/8MnO3 (LPCMO) [34]. Some anomalous metastable features associated with the AFM (charge ordered)-FM transition of this compound had actually been noticed earlier [35]. Commonality of the kinetic arrest observed in LPCMO and in other classes of magnetic materials has been discussed in details in Ref.5.

Before proceeding further, we wish to assert that the virgin curve will lie outside the envelope magnetization (M)- field (H) (or Resistance (R) – field (H)) curve if the measuring temperature falls below $T_K$(H=0), but also if it lies between T*(H=0) and T**(H=0)! In the latter case it matters whether the measuring temperature is reached by cooling or heating [11, 36,37]. It is important to note these in view of some recent confusion (see Fig. 3 in the reference [38]) and associated discussion). We now address the similarity mentioned in the preceding paragraph. For a cooling field lying between $H_1$ and $H_2$ (schematic-5), the equilibrium state and an arrested state coexist at the lowest temperature. Cooling in a higher field increases the FM fraction, irrespective of the equilibrium state; cooling in different fields and varying the field isothermally at the lowest T enables one to tune the fraction of the coexisting phases. If the state so obtained is warmed in a field different from the cooling field, then the equilibrium phase fraction can increase. For an FM (AFM) ground state, the equilibrium fraction increases if the warming field is larger (smaller) than the cooling field, and one sees a reentrant behavior during warming. This 'cooling and heating in unequal field' (CHUF) protocol not only allows the determination of the equilibrium state, but provides conclusive evidence of the occurrence of glass-like kinetic arrest of a first order phase transition. If various values of the cooling field are used in conjunction with a single warming (or measuring) field, then one observes two temperatures where nonergodicity sets in; one for $H_{Cool} > H_{Warm}$ and another for $H_{cool} < H_{warm}$. This is because there is a glasslike arrest, and also an underlying first order phase transition. The CHUF protocol has been very useful in identifying magnetic-glass behavior in many magnetic systems, and this is exemplified in Fig.7 with the results obtained in Co-doped NiMnSn magnetic shape memory alloys [39].

It is also important to highlight here a few experimentally determined characteristics of magnetic glass and the relevant set of experiments, which will enable one to distinguish a magnetic glass unequivocally from the well-known phenomena of spin glass (SG) and reentrant spin glass (RSG) [40]. First of all a magnetic-glass arises out of the kinetic arrest of a first-order FM to AFM phase transition, which is accompanied by a distinct thermal hysteresis between the FCC and FCW magnetizations, whereas no such thermal hysteresis is expected in the case of a SG/RSG transition since this is considered to be a second-order phase transition or a gradual phase transformation. Second, the thermomagnetic irreversibility associated with the magnetic-glass rises with the increase in the applied H, and this is just the opposite in SG/RSG. Third, the FC state in the magnetic-glass systems is the non-equilibrium state showing glass-like relaxation, whereas in the SG/RSG the ZFC state is the non-equilibrium state, which shows thermal relaxation. Lastly, the newly introduced experimental protocol CHUF reveals distinct features in the $T$ dependence of magnetization in magnetic-glass, which depends on the sign of inequality between the fields applied during cooling and heating; no such features are expected for a RSG system.

Both jamming and structural glass formation are manifestations of the slowing down of translational kinetics, and this leads to certain similarities between these two distinct phenomena. In the context of magnetic-glass, recently Chaddah and Banerjee [41] argued that the magnetic-glass formation arose due to kinetic arrest of the underlying first order phase transition, and not due to jamming. The phenomenon of jamming does not need any underlying first order phase transition or latent heat. The argument of Chaddah and Banerjee [41] is based on the idea that glasses are formed when the heat removal process preferentially removes specific heat, without removing latent heat. Within this picture magnetic glasses are likely to form in such systems where magnetic latent heat is weakly coupled to the thermal conduction process.

**3 Disorder influenced first order phase transition in ferroelectric materials**

The ferroelectric systems and ferromagnetic systems are quite similar in many ways. Below a critical temperature known as Curie temperature, electrical polarization and magnetic moment go to an ordered state in ferroelectric and ferromagnetic systems, respectively. In both the systems the order parameter can be coupled to lattice, and this in turn can lead to the piezoelectric/electro-strictive and magneto-elastic effects.

A first order phase transition can be induced both by temperature and electric field in the ferroelectric (FE) materials, leading to a sharp change in polarization accompanied with distinct thermal or electric field hysteresis. In many materials including Ba-doped $PbZrO_3$ [42], La-doped $PbZrO_3$[43], La and Hf doped $PbZrO_3$ [44], $Pb_{0.99}[(Zr_{0.8}Sn_{0.2})_{0.96}Ti_{0.04}]_{0.98}Nb_{0.02}O_3$ [45] and $PbHfO_3$ [46] there is an electric field induced first order phase transition from antiferroelectric (AFE) to ferroelectric (FE) state giving rise to a double hysteresis loop in the polarization (P) versus electric field (E) curve in the isothermal field excursion between $\pm E_{Max}$.

The role of disorder in the ferroelectric materials has also been investigated in some details. In one of the classic ferroelectric material $PbZrO_3$ the width of the temperature range in which the ferroelectric (FE) phase is stable, depends on the amount of defects in the $PbZrO_3$ [47,48]. It has also been observed that both FE and antiferroelectric (AFE) phases coexisted in epitaxial $PbZrO_3$ films with high crystalline quality [49]. Results of dielectric and x-ray diffraction measurements in $Pb_{0.90}Ba_{0.10}ZrO_3$ ceramics indicated that while cooling the AFE–FE phase transition in this material was not quite reversible [50]. It was observed that the equilibrium AFE phase could be recovered from the metastable FE matrix on ageing at room temperature, and the kinetics of recovery of this AFE phase was quite slow [50].

There is a class of ferroelectric materials called relaxor-ferroelectrics, which are distinguished from ordinary ferroelectrics by the presence of a diffused phase transition and strong metastable behaviour. They also show electric-field annealing and the aging effects, which indicate the evolution of micropolar clusters under an external electric field. The glass-like behavior in many relaxor-ferroelectrics Li- and Nb-doped $KTaO_3$, $PbMg_{1/3}Nb_{2/3}O_3$ and La-substituted $PbZr_{1-x}Ti_xO_3$ families are now quite well known [51]. While a detailed discussion of these materials is beyond the scope of the present paper, we will briefly discuss some phenomena in $KTaO_3$ which are of the interest in the present context. $KTaO_3$ is a quantum paraelectric where the ferroelectric transition is suppressed by zero-point fluctuations [52]. A small substitution of Li for K in $KTaO_3$ creates a local electric dipole due to its off-centre position with respect to the cubic site and a sharp ferroelectric phase transition has been observed in $K_{0.937}Li_{0.063}TaO_3$ [53]. A distinct thermal hysteresis and metastable behaviour associated with sharp peaks in the dielectric permittivity while cooling and heating in electric fields 75 kV m$^{-1}$ < E < 300 kV m$^{-1}$ clearly indicates the first order nature of this transition [54]. The nature of the low temperature phase of $K_{1-x}Li_xTaO_3$ (x <<1) is quite interesting, where signatures of both glass like and long-range ordered ferroelectric behaviour have been observed [55-58]. Similarities between the characteristic behavior of realaxor-ferroelectrics and two-phase coexistence, magnetic-field annealing and the aging effect, and diffuse x-ray scattering in various CMR-manganite systems have already been documented [59,60]. In fact Kimura et al highlighted that the "freezing of the two-phase coexistence state under the first-order phase transition accompanied by lattice distortion are common characteristics of CMR manganites and relaxor ferroelectrics" [59,60]. Such magnetic field annealing and metastable behaviors observed in the CMR-manganites can actually be explained quite naturally within the premise of the kinetic arrest of a first order magneto-structural phase transition [5]. This is brought out in some detail by Kumar et al [5] where detailed structure within the bands for supercooling and for kinetic arrest was also discussed. In this context various E–T history effects, metastability and glass-like behaviour in realaxor-ferroelectrics may be worth revisiting.

Signatures of the kinetic arrest of a first order phase transition and the associated glass-like metastable response have been observed in the multiferroic material $LuFe_2O_4$ [61-63]. This system undergoes a three-dimensional charge order transition below 320 K , which is followed by a ferrimagnetic transition below 240 K , and a first order magnetic phase transition coupled with monoclinic distortion takes at 175 K [61]. There is indication that this last magnetic transition is kinetically arrested giving rise to glass-like non-equilibrium behaviour [61-63]. Kinetic arrest (dearrest) of the first order multiferroic to weak-ferromagnetic transition has been reported in the multiferroic system $Eu_{0.75}Y_{0.25}MnO_3$, which results in frozen (melted) magneto-electric glass states with coexisting phases [64].

**4 Conclusions**

Whenever a first order phase transition occurs between two states with close-lying energies, it gives rise to a delicate balance of competing order parameters that are influenced by slight disorder, making the transition susceptible to interruption by glass-like kinetic arrest with consequent phase coexistence. Complex physical phenomena are also observed when there is a delicate balance of competing order parameters, as in multiferroic materials. Recent reports of kinetic arrest of the first-order multiferroic transition, resulting in magneto-electric glass states with two coexisting phases [64], point to the ubiquitous nature of the phenomenology of glasslike kinetic arrest. Questions related to the formation of structural glasses have persisted for over a century with the experimentalists having to study newer materials for testing newer ideas. The phenomenology of the kinetic arrest of first order phase transition discussed by us allows a second thermodynamic control variable, and its validity across magnetic and dielectric transitions may enable the resolution of outstanding issues in understanding glasses.

**Figures :**

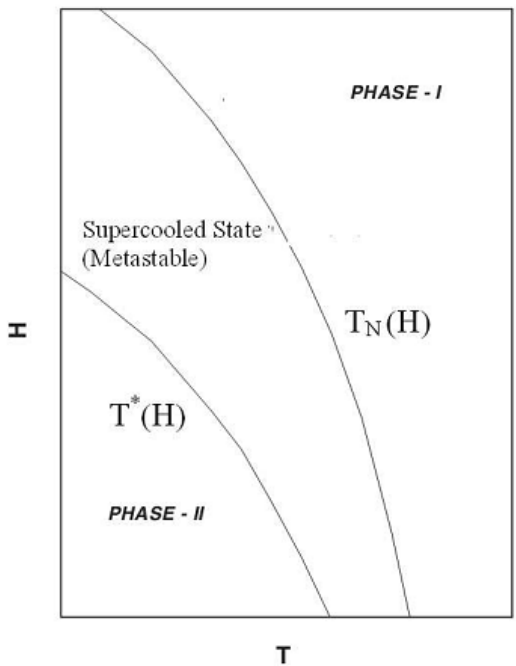

**Figure 1**. Schematic (H,T) phase diagram showing the phase transition line $T_N(H)$ and the limit of metastability $T^*(H)$ for the supercooled state [Ref. S B Roy, J. Phys.:Condens. Matter, **25** 183201 (2013)].

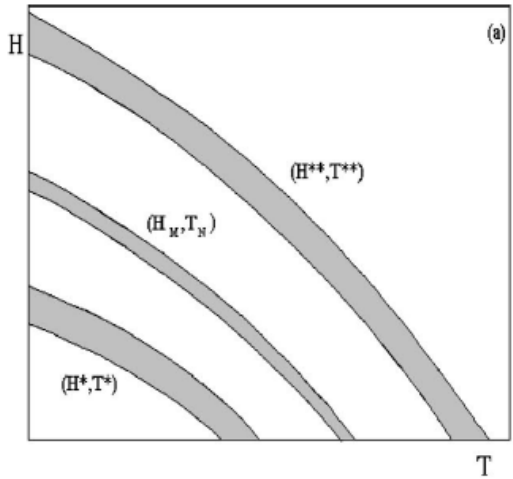

**Figure 2**. Schematic representation of broadened bands of phase-transition ($H_M, T_N$), supercooling ($H^*, T^*$), and superheating ($H^{**}, T^{**}$) lines [Ref. 25].

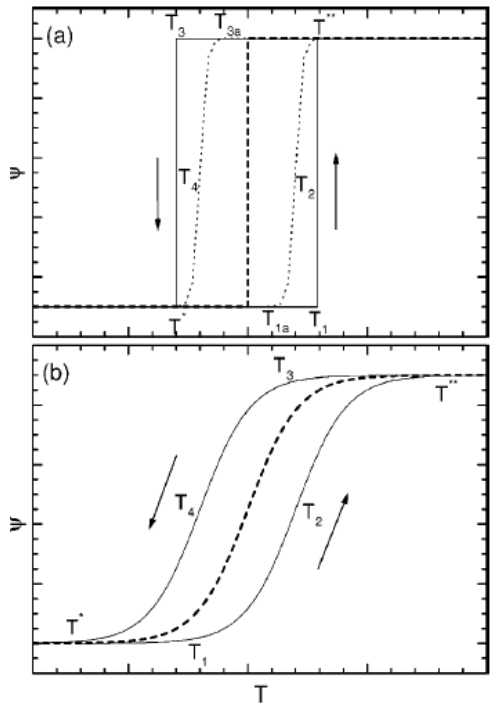

**Figure 3**. Schematic showing the fraction of transformed phase as a function of temperature in a first-order phase transition. (a) Sharp transition (b) Disorder-broadened transition [Ref. 26].

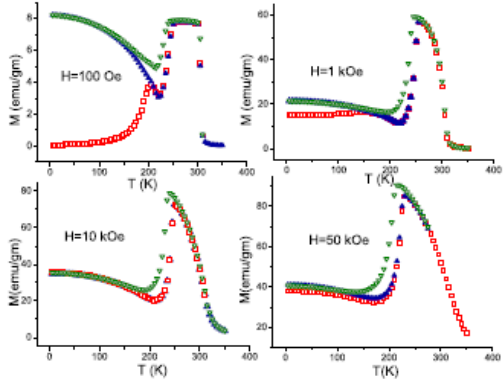

**Figure 4(a).** Magnetization (*M*) versus temperature (*T*) plots for $Ni_{50}Mn_{34}In_{16}$ alloy in external magnetic field *H*=0.1, 1, 10, and 50 kOe. Open square represents ZFC data and open (close) triangle represents FCC (FCW) data [Ref.7].

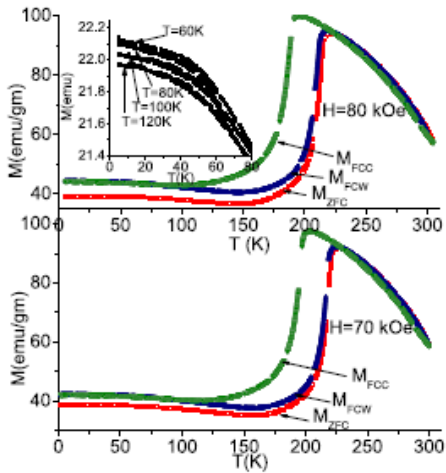

**Figure 4(b).** Magnetization (*M*) versus temperature (*T*) plots for $Ni_{50}Mn_{34}In_{16}$ alloy in external magnetic field *H*=70 and 80 kOe. The inset shows the effect of thermal cycling on $M_{FCC}$ (*T*), thus highlighting the non-equilibrium nature of the low T magnetic-glass state [Ref.7].

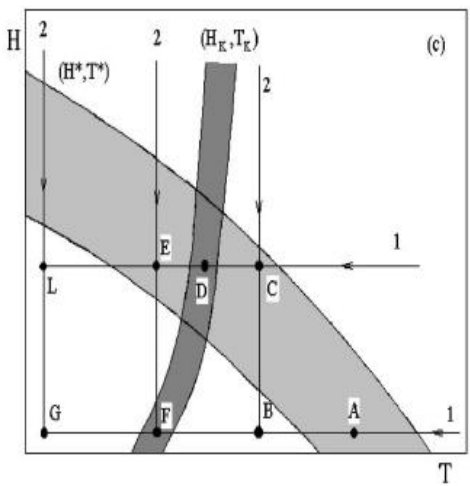

**Figure 5.** Schematic representation of the relative position of the band ($H_K$, $T_K$) across which the kinetics of the phase transition is hindered with respect to ($H^*$, $T^*$) band [Ref.25].

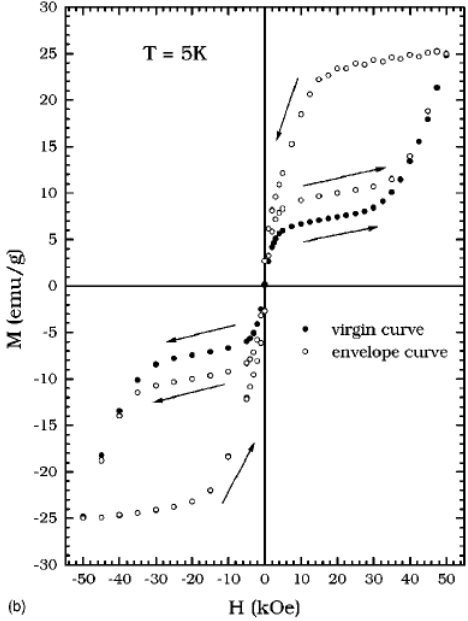

**Figure 6.** $M$ vs $H$ plots of Ce(Fe$_{0.96}$Al$_{0.04}$)$_2$ obtained after cooling in zero field at $T=5$ K. Note that the virgin $M$-$H$ curve lies outside the envelope $M$-$H$ curve. To confirm this anomalous nature of virgin curve, the same is also traced in the negative field direction after zero-field cooling the sample from the temperature above the magnetic phase transition [Ref.25].

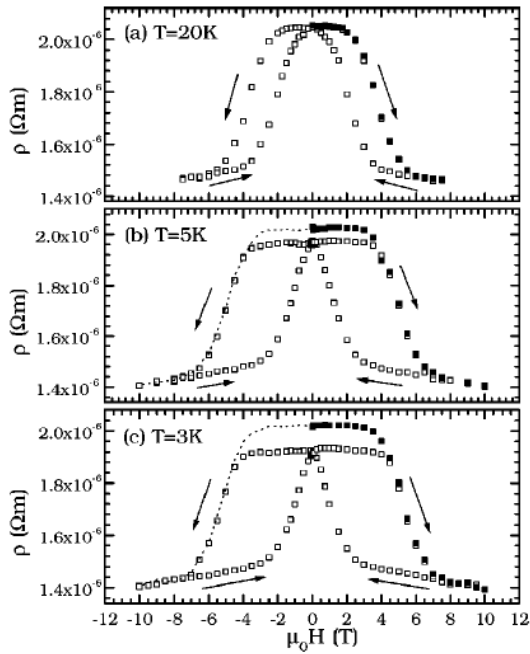

**Figure 7.** Resistivity ($\rho$) vs $H$ plots of Ce(Fe$_{0.96}$Al$_{0.04}$)$_2$ at $T = 20$ K, 5 K, and 3 K. Filled squares (dashed lines) represent virgin curve drawn in the positive (negative) field direction after zero-field cooling the sample [Ref.25].

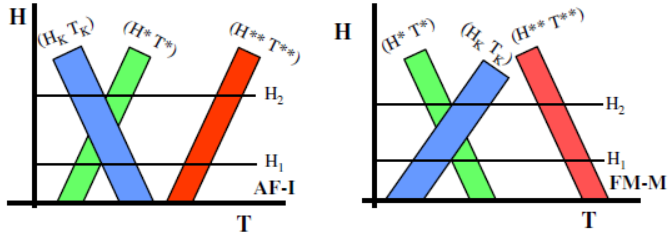

**Figure 8.** Comparison of schematic phase diagrams showing transformation from high-temperature AFM phase to low-temperature FM phase, and from high-temperature FM phase to low-temperature AFM phase [Ref.33].

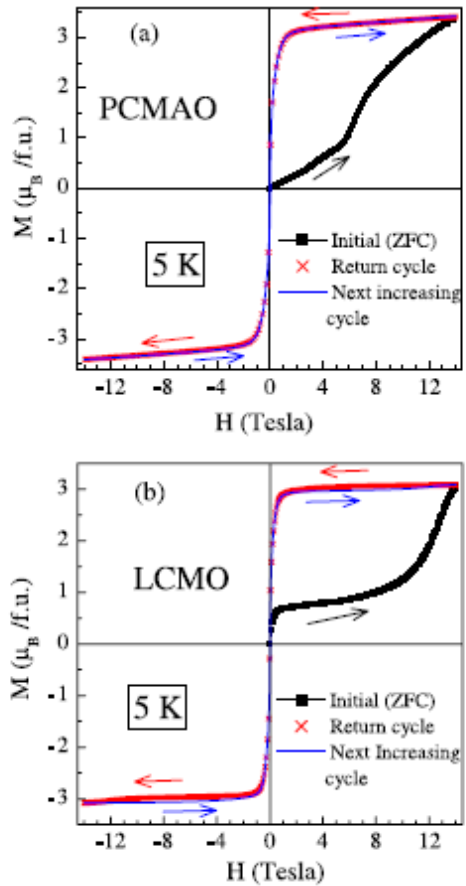

**Figure 9.** *M–H* curves of CMR manganite compounds PCMAO and LCMO at 5 K after cooling the samples from 320 K in zero field. Note that the the virgin M-H curve lies outside the envelope M-H curve in both the cases [Ref.10].

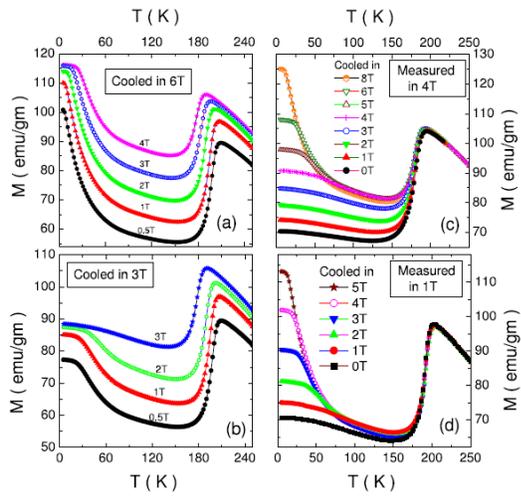

**Figure 10.** M vs T curves obtained using the CHUF (cooling and heating in unequal field) protocol in $Ni_{45}Co_5Mn_{38}Sn_{12}$ alloy. In (a), (b) the sample is cooled in a constant field of 6 or 3 T and measurements are carried out in various different fields. In (c), (d) the sample is cooled under different magnetic fields whereas measurements during warming are carried out in 4 or 1 T respectively [Ref.39].